\documentclass[journal,transmag]{IEEEtran}
\usepackage{amssymb}
\usepackage{bm}
\usepackage{booktabs}
\usepackage{threeparttable}
\usepackage{algorithmic}
\usepackage{multirow}
\usepackage{graphicx}
\usepackage{algorithm}
\usepackage{pifont}
\usepackage{amsmath}
\usepackage{xcolor,enumerate}
\usepackage{makecell}
\usepackage{multirow}
\usepackage{multicol}
\usepackage{enumitem}
\usepackage{wasysym}
\usepackage{framed}
\usepackage{pgfgantt,rotating}
\usepackage{subfloat}
\usepackage{multirow}
\usepackage{blindtext}
\usepackage[noadjust]{cite}
\usepackage{url}
\usepackage{algorithm,algorithmic}
\begin{document}
%
\title{Towards a Decentralized Digital Engineering Assets Marketplace: Empowered by Model-based Systems Engineering and Distributed Ledger Technology}


\author{\IEEEauthorblockN{Jinzhi Lu\IEEEauthorrefmark{1},~\IEEEmembership{CSEP},
Xiaochen Zheng\IEEEauthorrefmark{1},
Zhenchao Hu\IEEEauthorrefmark{2}, 
Huisheng Zhang\IEEEauthorrefmark{2}, and
Dimitris Kiritsis\IEEEauthorrefmark{1}}
\IEEEauthorblockA{\IEEEauthorrefmark{1} SCI STI DK,
Ecole Polytechnique Fédérale de Lausanne, Lausanne, 1015, Switzerland}
\IEEEauthorblockA{\IEEEauthorrefmark{2}School of Mechnical Engineering
Shanghai Jiaotong University, Shanghai, 200240, China} 
\thanks{Manuscript received December 1, 2012; revised August 26, 2015. 
Corresponding author: Xiaochen Zheng (email: xiaochen.zheng@epfl.ch).}}

\markboth{Journal of \LaTeX\ Class Files,~Vol.~14, No.~8, August~2015}%
{Shell \MakeLowercase{\textit{et al.}}: Bare Demo of IEEEtran.cls for IEEE Transactions on Magnetics Journals}
%

\IEEEtitleabstractindextext{%
\begin{abstract}
Model-based Systems Engineering (MBSE) has been widely utilized to formalize system artifacts and facilitate their development throughout the entire lifecycle. During complex system development, MBSE models need to be frequently exchanged across stakeholders. Concerns about data security and tampering using traditional data exchange approaches obstruct the construction of a reliable marketplace for digital assets. The emerging Distributed Ledger Technology (DLT), represented by blockchain, provides a novel solution for this purpose owing to its unique advantages such as tamper-resistant and decentralization. In this paper, we integrate MBSE approaches with DLT aiming to create a decentralized marketplace to facilitate the exchange of digital engineering assets (DEAs). We first define DEAs from perspectives of digital engineering objects, development processes and system architectures. Based on this definition, the Graph-Object-Property-Point-Role-Relationship (GOPPRR) approach is used to formalize the DEAs. Then we propose a framework of a decentralized DEAs marketplace and specify the requirements, based on which we select a Directed Acyclic Graph (DAG) structured DLT solution. As a proof-of-concept, a prototype of the proposed DEAs marketplace is developed and a case study is conducted to verify its feasibility. The experiment results demonstrate that the proposed marketplace facilitates free DEAs exchange with a high level of security, efficiency and decentralization. 
\end{abstract}

}

\maketitle

\IEEEdisplaynontitleabstractindextext
\IEEEpeerreviewmaketitle

\section{Introduction}

\IEEEPARstart{I}{n} the Industry 4.0 era, Internet of Things (IoT) and Cyber-Physical Systems (CPS) are constructed by multi\-domain physical compositions, networks, control units and computation components. These systems are considered as "system of systems" (SoS). The increasing functionalities bring new challenges for managing complex systems as every composition usually involves various stakeholders working together to develop the system from a SoS perspective \cite{Torngren2018}. During the entire lifecycle, co-design and collaborative design among stakeholders require frequent data exchange to gain insight to analyze, optimize and verify the system. For example, vehicle system engineers need to provide requirements to embedded system engineers in order to obtain the expected embedded systems. Thus, requirements and solutions need to be exchanged across organizations from the initial phase until the prototype is finalized. In these situations, identification of required "digital engineering assets (DEAs)" for each stakeholder is critical to the entire system development. A DEA includes documents, data and models used for information exchange. In different phases of a system lifecycle, different stakeholders may have different data and models as their digital assets. Currently it is still a challenging task to share related asset due to the lack of common understanding of expected digital asset and interoperability of heterogeneous data. The creation of an open and reliable marketplace can accelerate the DEAs exchange.

First, stakeholders exchange DEAs to realize information sharing thus to support co-design and collaborative design. It is not only difficult to identify the contents of stakeholders’ expected digital assets but also information represented by the digital assets must be correct, verifiable and unambiguous in different ways. In order to coordinate different stakeholders, systems engineering providing sets of standards and approaches are required to define the contents to construct DEAs \cite{Beihoff2010}. Moreover, aiming to develop digital asset correctly, model-based systems engineering (MBSE) is proposed to define the contents of digital assets using models \cite{Lu2017}. MBSE provides standardized specifications for constructing DEAs and defines formal descriptions for system artifacts of entire lifecycle.

Second, complex system development requires information exchange across organizations and lifecycle. Thus digital assets are expected by different stakeholders from different domains and hierarchies (requirement, function, behavior, etc.). However, each domain and hierarchy has specific semantics due to the fact that different domains use different mathematical theories, scopes and methods to implement their specific works \cite{Schluse2018}. Moreover, domain specifications often bring challenges to represent the entire systems using integrated syntax. These specific syntax and semantics lead to heterogeneous data structure which is a big challenge for data integration. Currently, meta-meta modeling approach is proposed to support data integration for DEAs \cite{DeLara2002}. This approach enables to construct domain-specific digital asset based on a high abstract level in order to realize the data integration during DEAs construction.

Third, the exchange of DEA among stakeholders requires an open and reliable marketplace. It is difficult to create such marketplace with traditional centralized approaches due to the concerns about data security and privacy. The stakeholders must trust each other or a third party during the DEAs exchange. It creates barriers when multiple partners are involved or the information need to be shared with partners from outside of the network. 
In recent years, the rapid development of decentralized Distributed Ledger Technologies (DLT), represented by the blockchain technology, provides an innovative tool to facilitate the concept of an open DEAs marketplace. A distributed ledger is considered as a distributed database maintained by a consensus protocol, which is run by nodes in a peer-to-peer network without any central administrator \cite{brogan2018}. As the most popular DLT protocol, blockchain was first applied to cryptocurrency systems like Bitcoin \cite{nakamoto2019bitcoin}. Because of its unique features, such as decentralized control, high anonymity and distributed consensus mechanisms, blockchain has gained attention from both academia and industry. It has been applied to variety of domains such as healthcare data sharing (\cite{zheng2019accelerating}), industrial IoT data exchange \cite{liu2018blockchain}, knowledge trading \cite{lin2019making} and energy dealing in smart grid (\cite{wang2019energy}) among others. The idea of using blockchain technology to empower the construction of data marketplace has also been proposed in some recent studies which mainly focused on health and industrial data (\cite{zheng2018blockchain,mikkelsen2018realization}).

This study aims to propose a decentralized open DEAs marketplace empowered by MBSE and DLT methodologies to cope with the challenges during DEAs exchange.
The rest of the paper is organized as follows. Related works are analyzed in Section \uppercase\expandafter{\romannumeral2} and the research methodology of this study is presented in Section 
\uppercase\expandafter{\romannumeral3}. The framework, key concepts and functional requirements of the proposed DEAs marketplace are presented in Section \uppercase\expandafter{\romannumeral4}. Enabling technologies, i.e. MBSE approaches and DLT solutions , are investigated in Section \uppercase\expandafter{\romannumeral5}. A proof-of-concept prototype of the marketplace is introduced based on a case study in Section \uppercase\expandafter{\romannumeral6}. Conclusions of this paper is presented in Section \uppercase\expandafter{\romannumeral7}. 

\section{Related works}
During digital asset construction and exchange, data interoperability is one challenge to stakeholders across organizations, because of heterogeneous data structure and intellectual property (\cite{Tavcar2019,Tao2019}). MBSE has been widely used to construct digital asset \cite{ Cloutier2015} and to support interoperability of the entire lifecycle of complex system development (\cite{ Yang2014,Mordecai2018}). It refers to a systems engineering approach focusing on creating and exploiting domain knowledge as the primary means of information exchange between stakeholders based on models during the entire lifecycle \cite{Ramos2012}. In Chen \textit{et al.}'s survey, we find that in the lifecycle of digital asset, the existence of various
data format standards is the basis to construct digital asset. MBSE approaches, such as SysML \cite{Cloutier2015}, can offer the defined specifications to formalize the system artifacts. Moreover, digital thread is a main approach to manage interoperability between digital twins and traceability between system development process and digital twin assets, because MBSE provides a unified descriptions of multiple domain knowledge and a specification to formalize system architecture views for the complex systems \cite{Singh2018}. Furthermore, the unified specification enables stakeholders to develop their own models and provide them with the information in black box. This is a proper way to protect their IP and to support information exchange during system co-design (\cite{Do2014,VanZandt2015}).

The utilization of DLT in data exchange or asset trading systems could enable better control and fine-grained tracking of different usages. Liu \textit{et al.} \cite{liu2018blockchain} proposed a data collection and exchange scheme, which utilized Ethereum blockchain technology to ensure security and reliability during industrial IoT data sharing. Aiming at solving the interoperability and trust issues for IoT services, Wattana \textit{et al.} \cite{viriyasitavat2019new} integrated blockchain technology with a service-oriented architecture to assure data validity, quality of services and uncertainties. Beyond the data sharing, in \cite{lin2019making} the authors developed a consortium blockchain to facilitate knowledge management and trading based on edge computing devices and edge artificial intelligence. The architecture of a peer-to-peer (P2P) knowledge market was proposed and a pricing strategy with incentives for the market based on noncooperative game theory was also discussed in this study. 
The idea of creating a distributed marketplace using blockchain technology has been proposed recently to remove the centralized components. In \cite{mikkelsen2018realization}, 
the authors used a private Ethereum blockchain to support the construction of an IoT marketplace making functionalities and storage of the marketplace distributed. In order to accelerate the IoT and other open data sharing, Gowri \textit{et al.} \cite{ramachandran2018towards} proposed a decentralized data marketplace based on blockchain to replace the existing centralized data marketplaces.

As a relatively new concept, decentralized marketplace enabled by blockchain has not been well studied and requires more research efforts. After reviewing existing studies, we found several gaps: 
\begin{itemize}
    \item Most existing marketplace solutions only focus on the exchange process of digital assets, while the interoperability of the assets remains as a challenge. We believe that data interoperability should be taken into account when designing a marketplace and MBSE could be a solution for this issue.
    \item There are few, if not no, studies focusing on the marketplace for supporting model-based systems development, which is usually more complex with multiple stakeholders.
    \item Due to the scalability limitations of most public blockchain protocols, most decentralized marketplace solutions utilize permissioned, i.e. private and consortium, blockchain protocols. These solutions are not feasible in most system development scenarios where unknown and trustless stakeholders are involved.
    \item Due to the high cost of blockchin transactions, most blockchain-based marketplaces are not suitable for big data exchange, while large size datasets are common during systems development.     
\end{itemize}
Considering these gaps, this paper aims to propose a novel decentralized marketplace solution by combing advanced MBSE and DLT approaches.

\section{Research Methodology}

\subsection{Systems thinking}

Systems thinking is an approach for understanding systems by examining the interactions between the components within the system boundary \cite{Haskins2014}. In this paper, systems thinking is adopted for designing our approach in order to provide a complete solution of the DEA marketplace based on the identified challenges and motivations .

\begin{itemize}
\item \textit{\textbf{Identify the scenarios of DEA marketplace:}} The scenarios are defined to understand the scope (systems boundary) and the related concepts about DEAs. For example, in the case study, one scenario including requirement distributions and solution receptions is defined. It demonstrates the processes that vehicle systems engineers distribute their requirements of the embedded systems and they get the solutions from the embedded system suppliers.
\item \textit{\textbf{Define the entities related to the scenario:}} In the scenario, the entities are captured, such as the required \textit{models} for describing expected \textit{views} of DEAs. 
\item \textit{\textbf{Define the interrelationships between entities:}} Interrelationships between entities refer to relationships between entities, for example, the inclusions between views and models \cite{Standard2011}. 
\item \textit{\textbf{Develop an approach for constructing DEA marketplace:}} Based on the entities and their interrelationships, an approach based on MBSE and DLT is proposed to construct the asset marketplace.
\item \textit{\textbf{Instantiate the entities using a case study:}} Based on the designed scenario, a case study is proposed for developing an approach for DEA exchange. 
\item \textit{\textbf{Evaluation of the case study:}} Based on the case study, qualitative and quantitative analysis is conducted for evaluating the proposed approach.
\end{itemize}


\section{DEAs marketplace concept, framework and requirements}
To facilitate the development of the DEAs marketplace, it is critical to formally define concept of DEAs and specify the functional requirements of the marketplace. This section firstly presents a conceptual framework of the proposed DEAs marketplace; then formally defines the concept of DEAs; and specify the functional requirements.



\subsection{DEAs marketplace framework}
\begin{figure}[!t]
\centering
\includegraphics[width=3.2in]{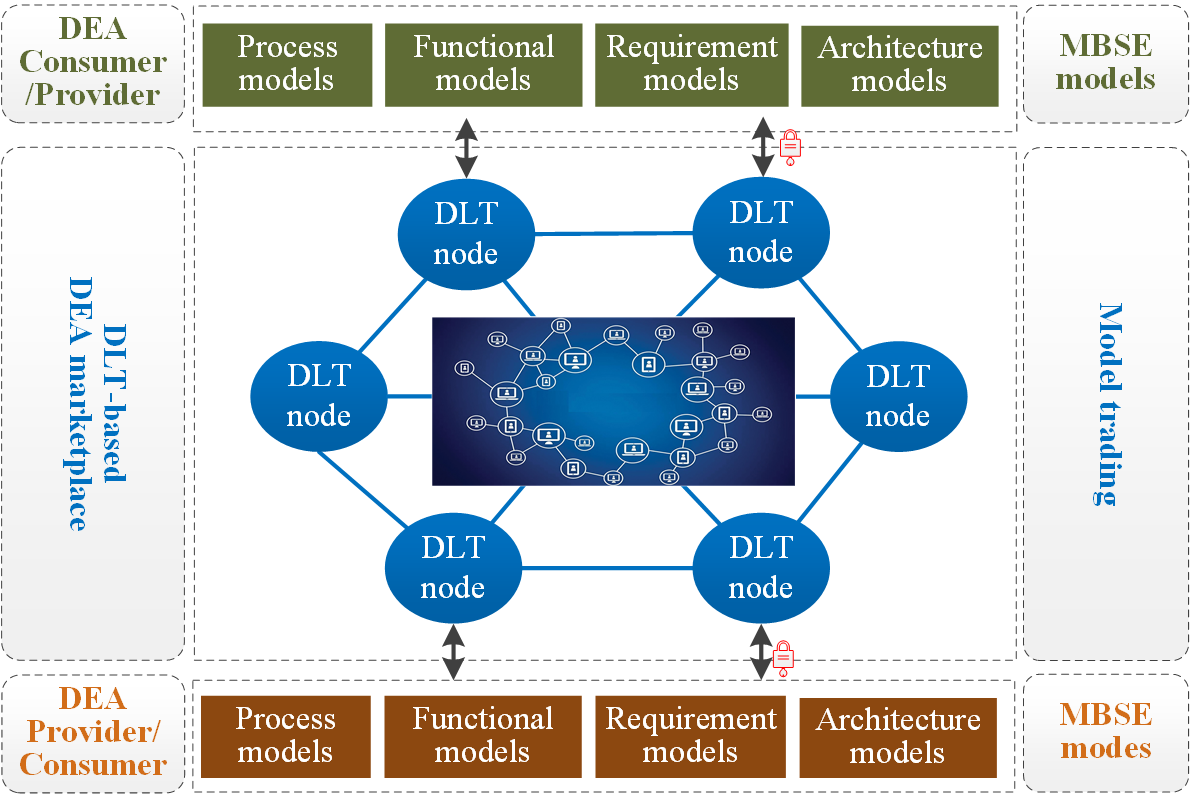}
\caption{General architecture of the proposed DEA marketplace}
\label{F:Unifo}
\end{figure}

The conceptual framework of the proposed DEAs marketplace is as shown in Fig. \ref{F:Unifo}. It is composed with four basic elements including DEAs which are the "goods" to be traded, DEA providers, DEA consumers, and the decentralized marketplace. 
DEAs providers formalize their development processes, requirements, functions and architectures following MBSE methodology to create DEAs. DEAs (or the meta data of DEAs) are added to the  marketplace through a public or private node which is connected to the marketplace network. DEAs consumers find interested DEAs through proactive searching or passive subscription. Consumers receive the DEAs directly or through certain authorization procedures depending on the trading mode.


\subsection{Definition of DEAs}

\begin{figure}[t]
\centering
\includegraphics[width=3.5in]{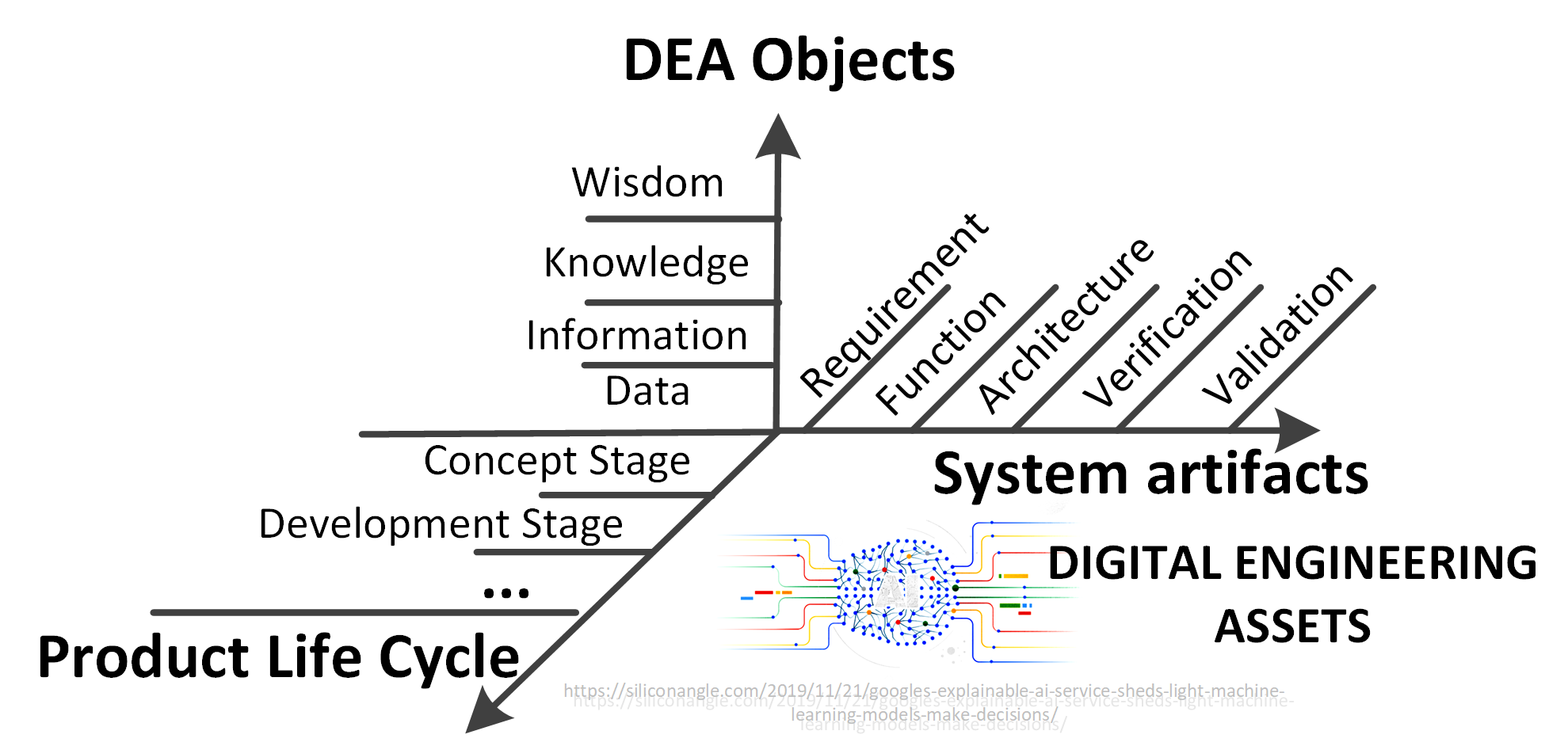}
\caption{Three Dimensional DEA}
\label{F:Vset}
\end{figure}

A DEA refers to a currency format in some specific case which the currency can be considered as either a medium of information exchange or an attribute that has value during the entire life cycle, such as virtual models or documents \cite{Styhre2003}. In this paper, a three dimensional DEA concept model is proposed to define the DEA from three main viewpoints: 1) DEA object; 2) product lifecycle; 3) system artifacts of products. First, the DIKW model is used, with roots in knowledge management, to classify the DEA objects from Data (D) to Information (I), Knowledge (K) and Wisdom (W) \cite{Lu2018}. It is used to represent the interrelationships between currency used during the entire lifecycle. The details are introduced as follows:

\begin{itemize}
\item Dimension of the product processes refers to the development process including various stages, phases and work tasks.
\item Dimension of system artifacts refers to views of system stakeholders, e.g. requirement, function and architecture.
\item Dimension of DEA objects refers to attributes of items including data, information, knowledge and wisdom \cite{Rowley2007}.
\begin{itemize}
\item 	Data items refer to attributes of DEAs implementing simulations, design, analysis, verification \& validation and optimization in the entire lifecycle. They are functional symbols, but not structure model or documents.
\item 	Information items refer to structural contents constructed by data, such as SysML models for describing system requirements.
\item 	Knowledge items refer to validated information which makes possible the transformation of DEAs into instructions.
\item 	Wisdom items refer to digital assets to support decision-makings based on knowledge, such as machine learning models for decision-making at one decision gate.
\end{itemize}
\end{itemize}

\newtheorem{definition}{Definition}
\begin{definition}
Token \textit{::=} refers to a collection of concepts. \textit{$DeaSs_{sys}$} refers to a collection of DEAs of a system \textit{sys}. 
\end{definition}

\begin{equation} 
\begin{aligned} 
\textit{DeaSs}_{sys} ::= \sum{\textit{$DeaS_{sys}^{(t, view, \alpha)}(i)$} }
\end{aligned}
\end{equation}
where \textit{$DeaS_{sys}$} refers to one DEA \textit{i} of \textit{sys}; \textit{$t$} refers to the time stamp of the related DEA during development process; \textit{$\alpha$} refers to the DIKW attributes of the digital asset.

\subsection{Functional requirements of the DEA marketplace}
Various DLT protocols and solutions have been developed in recent years with different advantages and limitations. It is critical to choose the proper solution according to the requirements of DEA exchanges, which cover the whole lifecycle of a system and may involve stakeholders from both inside and outside of an existing organization or alliance network. Based on practical industrial needs and previous studies (\cite{ramachandran2018towards,lin2019making}) we define the functional requirements of the proposed DEA marketplace as follows.

\begin{itemize}
    \item \textbf{Decentralization:} Stakeholders of the marketplace are usually geographically distributed and might have no trust with each other. Therefore, the digital assets should be managed in a decentralized way without any dominant party. All stakeholders should be enabled to join the marketplace fairly and freely. 
    \item \textbf{Nontampering:} The marketplace should be able to protect the DEAs from the malicious attacks to keep the integrity and accuracy. Besides, the trading process and records should also be protected to avoid tampering. 
    \item \textbf{Flexible access control:} The marketplace should provide different DEA access authorization approaches for different scenarios. For example, a stakeholder's required DEA might need to be accessible to all stakeholders to find as many potential suppliers as possible; in contrast, a supplier's proposal DEA should be open only to the customer and relevant parties.  
    \item \textbf{Scalability:} With the expanding of the marketplace the number of stakeholders and trading frequencies will increase rapidly. The marketplace should enable high scalability with large, even unlimited, transaction throughput and low confirmation delay. 
    \item \textbf{Low transaction cost:} The value of a DEA varies in the marketplace. There might be several iterations during a development lifecycle. The cost of a transaction should be as low as possible. It will make no sense if the transaction fees become close or higher than the DEA per se. 
    \item \textbf{Big data support:} A DEA might contain large size documents like product models. The marketplace should support the exchange of large files with short delays. 
    \item \textbf{Interoperability:} The DEAs of different stakeholders usually come with high heterogeneity, e.g. different formats, sizes and protocols. The marketplace should facilitate the interoperations among different stakeholders. 
\end{itemize}

\section{Enabling technologies}

\subsection{GOPPRR approach}

\begin{figure}[!t]
\centering
\includegraphics[width=3.5in]{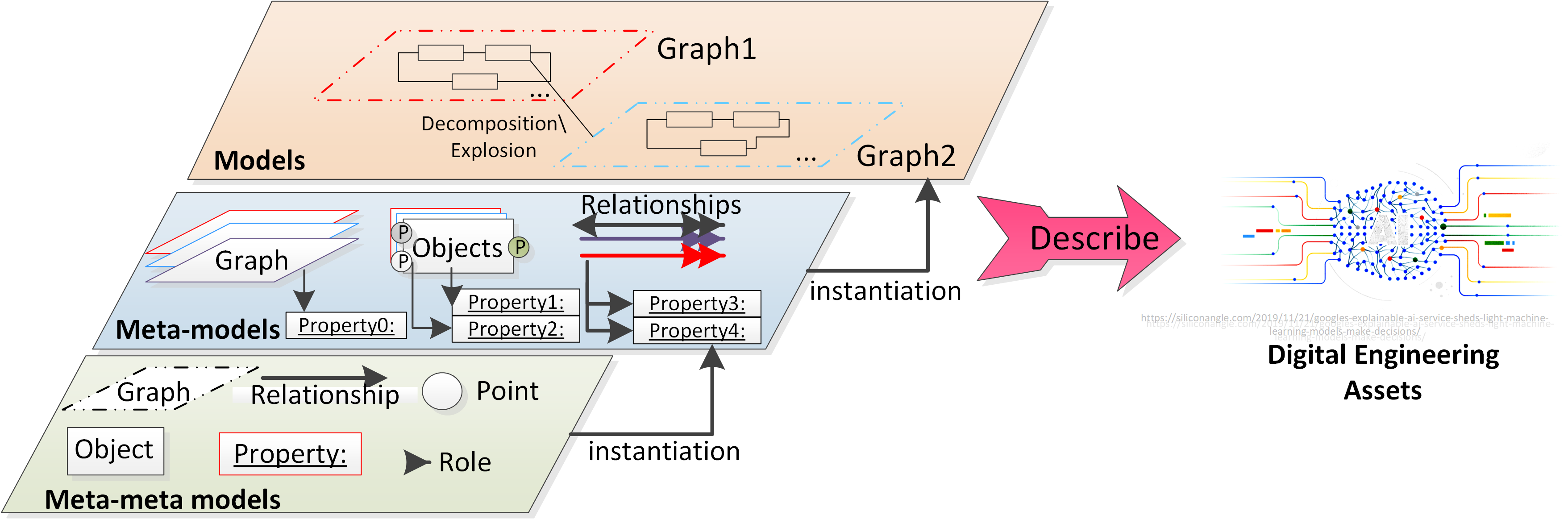}
\caption{GOPPRR Formalizing DEA marketplace}
\label{F:GOPPRR}
\end{figure}

This section introduces a GOPPRR approach for formalizing the DEA on the three-dimensional conceptual model in Section  IV.B \cite{Lu2018}. It is adopted to develop meta-models for constructing the MBSE models aiming to describe DEAs from system artifacts, development processes and DEA objects.  The \textit{GOPPRR} approach is one the most powerful approaches to describe domain specific characteristics and meta-meta models of products \cite{Kern2011}, including \textbf{G}raph, \textbf{O}bject, \textbf{P}oint, \textbf{P}roperty, \textbf{R}ole and \textbf{R}elationship: 
\begin{itemize}
\item \textit{\textbf{G}raph} is a collection of \textit{Object}, \textit{Relationship} and \textit{Role} represented as one window (one integrated concept of a class diagram and package in UML). The graph can be a visual diagram on the top level or lower level decomposed by one \textit{Object}.
\item \textit{\textbf{O}bject} is one entity in \textit{Graphs} (class concepts in UML).
\item \textit{\textbf{P}oint} is a port in \textit{Objects}.
\item \textit{\textbf{R}elationship} is one connection between the different \textit{Points} of \textit{Objects}.
\item \textit{\textbf{R}ole} is used to define the connection rules mirrored to the relevant \textit{Relationship}. For example, one \textit{Relationship} has two \textit{Roles}. Each is defined to connect with one \textit{Point} in \textit{Objects}. Then, the \textit{Relationship} is connected with these \textit{Points} in the \textit{Objects}.
\item \textit{\textbf{P}roperty} refers to one attribute of the other five meta-meta models.
\end{itemize}

\begin{definition}
 $\mathit{\textit{Graph}_{Gi}^a}$ refers to the model \textit{a} based on the meta-model of Graph \textit{Gi};In  $\mathit{\textit{Graph}_{Gi}^a}$,  $\mathit{\textit{Object}_{Obji}^b}$ refers to the object instance \textit{b} based on the meta-model of Object \textit{Obji}; $\mathit{\textit{Relationship}_{Rei}^c}$ refers to the relationship instance \textit{c} based on the meta-model of Relationship \textit{Rei}; $\mathit{\textit{Role}_{Roi}^d(x)}$ refers to the role instance \textit{d} based on the meta-model of Role \textit{Roi} in the meta-model Relationship \textit{Rei}; $\mathit{\textit{Point}_{Poi}^e(y)}$ refers to the point instance \textit{e} based on the meta-model of Point \textit{Poi} in the meta-model Object \textit{Obji}; $\mathit{\textit{Property}_{Proi}^f(x)}$ refers to the property instance \textit{Proi} based on the meta-model of Property \textit{f} in the meta-model \textit{z} ($z \subseteq \{Gi, Obji, Rei, Roi, Poi\}$);
\end{definition}

\begin{equation} 
\begin{aligned} 
\textit{Graph}_{Gi}^a ::= ( \sum\textit{Object}_{Obji}^b, \sum\textit{Relationship}_{Rei}^c,\\ \sum\textit{Role}_{Roi}^d(x), \sum\textit{Point}_{Poi}^e(y), \sum\textit{Property}_{Proi}^f(z))
\end{aligned}
\end{equation}
where each Relationship \textit{c} is bond to two Objects or Points in Objects through Roles in order to identify the connections among Objects and Points.

\begin{definition}
Token $\mathit{\Rightarrow}$ refers to the realizations between MBSE models and DEAs.
\end{definition}

\begin{equation} 
\begin{aligned} 
\textit{Graph}_{Gi}^a \Rightarrow \textit{$DeaS_{sys}^{(t, view, \alpha)}(a)$}
\end{aligned}
\end{equation}
where model $\textit{a}$ formalizes the DEA \textit{i}.


\subsection{DLT solutions}

To find the most feasible DLT solution for the proposed marketplace, we firstly compared different DLT solutions on the category level. In general, as shown in Table \ref{table:function}, DLT can be classified into public and permissioned solutions. Depending on the permission governance scheme, permissioned DLT can be further divided into private and consortium solutions. Public DLT allows any stakeholder that holds a valid pseudonym to make and verify transactions in a secured and trustless environment, and even set and connect their own nodes to the network without requesting authorization from any party. In contrast, the permissioned DLT platforms are controlled by a single entity or a number of pre-selected entities corresponding to private and consortium solutions.

In general, most public DLT solutions have high security and decentralization, but usually have relatively low scalability and high transaction cost. In comparison, permissioned DLT solved the scalability problem by reducing decentralization. As the proposed DEA marketplace should be open to all participants and have no dominant entities, permissioned solutions will not fulfill the requirements. Therefore, in this study we only focus on public DLT solutions.

\begin{table}[]
\begin{center}
\caption{Comparison of DLT categories}
\label{table:function}
\begin{tabular}{lccc}\toprule
\multirow{2}{*}{DLT category} & Permissionless   & \multicolumn{2}{c}{Permissioned} \\\cline{2-4} 
                 & Public          & Consortium   & Private    \\  \hline
Decentralization & \checkmark      & \checkmark   &  \ding{55} \\
High security    & \checkmark      &  \ding{55}   & \ding{55}  \\
Free access      & \checkmark      & \ding{55}    & \ding{55} \\
High throughput  & Depends on protocol   & \checkmark   & \checkmark \\
Low cost         & Depends on protocol       & \checkmark &  \checkmark  \\ \hline
\end{tabular}
\end{center}
\end{table}

\begin{table*}[]
\begin{center}
\caption{Comparison of popular public DLT solutions}
\label{table:dlt}
\begin{tabular}{lllllll}\toprule
\multirow{2}{*}{Type} & \multirow{2}{*}{DLT solutions} & \multirow{2}{*}{Protocol} & \multicolumn{4}{c}{Properties}  \\ \cline{4-7} 
 &   &  & Throughput (tx/s) & Confirmation time & Fault tolerance  & Transaction fees\\ \hline
\multirow{4}{*}{Blockchain-based} & Bitcoin  & PoW   & Tens & $\approx 60 $ min.& 50\% power    & $ \geq 0.3 $ USD \\
& Ethereum & PoW   & Tens        & $\approx 6 $ min. & 50\% computing power    & $ \geq 0.1$ USD   \\
& Waves-NG & PoW (key blocks)  & Hundreds & $\approx 10 $ min. & 50\% computing power    &  0.003 Waves \\
& Litecoin & PoW     & Tens        & $\approx 30 $ min. & 50\% computing power  &  0.001 LTC \\
\multirow{2}{*}{DAG-based} & IOTA   & PoW    & $\geq $ Thousands & 0.5 to 10 min.   & 50\% computing power  & None    \\ 
& Nano   & PoW     & $\geq $ Thousands  & 0.5 to 10 min.   & 50\%  token wealth  & None  \\
\multirow{2}{*}{Other protocols} & Qtum     & PoS     & Thousands  & $\approx 60 $ min. & 50\% stake value  & 0.001 Qtum/KB \\ 
& Cardano  & PoS       & Tens     & $\approx 10 $ min.& 50\% stake value  & $\geq 0.16$ ADA \\ \hline
\end{tabular}
\end{center}
\end{table*}


We reviewed some of the popular public DLT solutions and evaluated their performances according to the aforementioned functional requirements of the DEA marketplace as shown in Table \ref{table:dlt}. The results show that, according to the consensus protocols, most public DLT can be classified into three categories, i.e. blockchain-based, DAG-based and other types \cite{xiao2020survey}. The most adopted consensus protocols behind blockchain-based public DLTs are the Nakamoto protocol \cite{nakamoto2019bitcoin} and its improved versions such as Nakamoto-GHOST \cite{sompolinsky2015secure} and Bitcoin-NG \cite{eyal2016bitcoin}. 
However, with the rapid increasing of the networks size, several drawbacks and vulnerabilities of Nakamoto protocol have arisen: 1) low transaction rate and poor scalability; 2) high transaction fees and inefficient energy consumption; 3) centralization risks due to mining pools; 4) vulnerable to quantum attack. For example, the whole Bitcoin network, with approximately ten thousands nodes in total, can approve less than seven transactions per second and each transaction costs around 0.3 USD according to the latest statistics (cite{fees2020}. In terms of computing power, the top five mining pools control more than 60\% of the whole network’s mining power \cite{pools2020}.

To overcome the drawbacks of the blockchain-based DLTs, several new protocols based on non-linear ledger structures have been proposed, among which Directed Acyclic Graph (DAG) has attracted much attention. Compared with blockchain-based protocols, the main advantage of DAG-based protocols is that they eliminate transaction throughput caps. Instead of sequentially storing all transactions to a linearly growing chain of blocks with fixed time intervals, in a DAG-based network every vertex can provide multiple diverging branches. Theoretically, a DAG-based protocol could allow millions of even unlimited transaction throughput, although in reality the capacity will be capped by physical communication bandwidth. These unique characteristics make DAG-based ledger structures the most favorable solutions for the construction of a open marketplace as required in this study.

IOTA Tangle is a DAG-based protocol specifically designed to support industrial IoT data exchange and machine-to-machine micro transactions \cite{popov2016tangle}. To issue a new transaction in the Tangle, users must perform a small amount of computational work to validate two unapproved transactions (tips), and this new transaction will be validated by some subsequent transactions. According to this two-tip rule, every vertex in the Tangle has an out degree of two and contains a Proof-of-Work (PoW). By this way, the scalability issue can be solved as the more transactions added to the Tangle, the faster transactions can be validated. Besides, this 'pay-it-forward' system of validations can be used as the incentive mechanism for honest participation, which eliminates financial rewards and makes transactions fee-less. Moreover, there are no miners thus no mining pools in the Tangle which makes it possible to be truly decentralized. 

Another advantage of IOTA Tangle is the provided second-layer data communication protocol named Masked Authenticated Messaging (MAM). It supports publishing and receiving encrypted data over the Tangle regardless of the size or cost of a device \cite{mam2017}. MAM distributes data using the concept of channels supported by the Tangle gossiping propagation mechanism. Any user who want to publish data can create a channel with a unique address and then attach data to this channel following the Tangle transaction two-tip approval procedure. Once approved, the other users who subscribe to this channel will be able to receive the data which might be encrypted according to the publisher's configuration. 

MAM uses Merkle Hash Tree (MHT) as the signature scheme to encrypt the message \cite{mam2017}. The root of a MHT is created using the unique user identification and it serves as the address of the data channel. Enabled by the MHT signature scheme, MAM allows different privacy and encryption modes to control the visibility of a channel and the access to the data i.e. public and restricted modes. Public mode uses the MHT root as both channel address and transaction identification. In this mode, any user who knows the channel address, even randomly, can decode and consume the message. 
Restricted mode adopts an authorization key based on private mode. The channel address is encrypted using the hash of the MHT root and the message with authorization key. This mode allows a publisher to revoke access to future messages by changing the authorization key without changing the channel address. This encryption mechanism fulfills the flexible access control requirement of the proposed DEA marketplace. 

Theoretically, there is no limitation about the size of a MAM message. When the size of a MAM message exceeds the transaction capacity, which is 2187 trytes (1.3KB), it will be automatically split into multiple transactions which will be contained in the same bundle to keep them retrievable. In the ideal scenario, i.e. when the IOTA Tangle network is mature enough to support millions or unlimited TPS, the exchange of large files will not be a problem. Currently, the IOTA Tangle is still under development with limited transaction processing capability, thus large MAM messages might cause delay in confirmation. 
As a temporary mitigation solution, the InterPlanetary File System (IPFS) \cite{benet2014ipfs} is adopted in this study to handle large DEAs. IPFS is a content-addressing, peer-to-peer network for storing and sharing arbitrary data in a distributed file system. It uses a hash to access a IPFS file and the hash value changes when any change is made to the content. Therefore, a large size DEA in the marketplace can be saved in the IPFS network and only its hash is traded in the marketplace. Before saving to IPFS, the content can be encrypted to assure the privacy and security either using the encryption mechanism provided by MAM or extra encryption algorithms such as AES256-GCM \cite{nechvatal2001report}.

In summarize, in order to address the functional requirements of the DEA marketplace,
MBSE tools, e.g. GOPPRR, are used to facilitate the interoperability among stakeholders by creating formalized DEAs; IOTA Tangle and MAM protocol are adopted to enable decentralization, nontampering, high scalability, low transaction cost and flexible access control requirements; and IPFS is proposed as a temporary solution to handle large files before the mature of IOTA Tangle network. 

Based on the selected enabling technologies, the overall workflow of creating and publishing DEAs is defined as shown in Algorithm \ref{alg1}. The receiving and consuming process is in the opposite sequence of the publishing process. Both processes are demonstrated through a case study in the following section.

\begin{algorithm}
 \caption{Publishing DEAs to the marketplace}
 \label{alg1}
 \begin{algorithmic}[1]
 \renewcommand{\algorithmicrequire}{\textbf{Input:}}
 \renewcommand{\algorithmicensure}{\textbf{Output:}}
 \REQUIRE $ \sum\textit{Object}_{Obji}^b, \sum\textit{Relationship}_{Rei}^c, \sum\textit{Role}_{Roi}^d(x),$ \par
  \hskip\algorithmicindent $\sum\textit{Point}_{Poi}^e(y), \sum\textit{Property}_{Proi}^f(z) $
 \ENSURE  IOTA Tangle transaction address
 \\ \textbf{Initialisation} : Create MBSE models: \par
 \hskip\algorithmicindent $\textit{Graph}_{Gi}^a ::= ( \sum\textit{Object}_{Obji}^b,  \sum\textit{Relationship}_{Rei}^c,  $ \par
 \hskip\algorithmicindent $\sum\textit{Role}_{Roi}^d(x),\sum\textit{Point}_{Poi}^e(y), \sum\textit{Property}_{Proi}^f(z)); $ \par
 \hskip\algorithmicindent Realize the DEAs from the MBSE models:\par
 \hskip\algorithmicindent $\textit{Graph}_{Gi}^a \Rightarrow \textit{DeaS}_{sys}^{(t, view, \alpha)}(a)$ 
 \FOR {$DeaS_{sys}^{(t, view, \alpha)}(a)$ from $DeaSs_{sys}$ }
  \STATE Encode DEA to IOTA Trytes:\par
  \hskip\algorithmicindent $DeaS_{sys}^{(t, view, \alpha)}(a) \Rightarrow \textit{Trytes}s_{dea(a)}$ 
  \STATE Encrypt $Trytes$ with MAM signature scheme:\par
  \hskip\algorithmicindent $\textit{Trytes}_{dea(a)} \Rightarrow (\textit{Encrypted}_{dea(a)},\textit{root}_{dea(a)})$
    \STATE Upload $\textit{Encrypted}_{dea(a)}$ to IPFS
    \RETURN IPFS content address $\textit{Hash}_{dea(a)}$ 
    \STATE Encode to IOTA $Trytes$: \par
    \hskip\algorithmicindent $(\textit{Hash}_{dea(a)},\textit{root}_{dea(a)}) \Rightarrow \textit{Trytes}_{Hash(a)}$ 
    \STATE Encryption with MAM signature scheme: \par
    \hskip\algorithmicindent $\textit{Trytes}_{Hash(a)} \Rightarrow (\textit{Encrypted}_{Hash(a)}, \textit{root}_{hash(a)})$ 
    \STATE Publish $\textit{Encrypted}_{Hash(a)}$ to IOTA Tangle by MAM
    \IF {\textit{public} mode}
      \STATE $\textit{Address}_{Hash(a)} = \textit{root}_{hash(a)} $
    \ELSIF{\textit{restricted} mode}  
      \STATE $\textit{Address}_{Hash(a)} = \textit{Hash}( \textit{root}_{hash(a)}) $
    \ENDIF
    \RETURN IOTA transaction address $\textit{Address}_{Hash(a)}$  
 \ENDFOR
 \end{algorithmic}
 \end{algorithm}

\section{Case Study}

The case study is based on a common scenario during vehicle system development.
An assembling vehicle company expects to purchase a solution of embedded system for a new vehicle model from external suppliers. The vehicle systems engineers first develop a requirement model, i.e. as the DEAs for the marketplace, using the GOPPRR approach. In this case, this message is expected to reach as many stakeholders as possible. Therefore, this DEA is published to the marketplace with public mode so that anyone can receive and consume the requirement model.

On the other side, the embedded system engineers from a supply agency receive the requirement model from the marketplace. Based on requirements, they develop solutions and build the solution models using the GOPPRR approach. In this case, the solution models are encrypted so that they are accessible only to authorized parties, i.e. the assembling vehicle company. Therefore, these DEAs (solution models) are published to the marketplace with restricted mode. The vehicle systems engineers receive the solution models from the marketplace and the corresponding encryption keys from supply agencies, so that they can access to the solutions and compare them to find the most suitable solution.

\subsubsection{DEAs marketplace prototype}
A prototype of the proposed DEA marketplace was constructed as proof-of-concept based on the IOTA Tangle API (\textit{iota.lib.js}), MAM API (\textit{mam.client.js}) and IPFS API (\textit{ipfs} Javascript implementation). The source codes, written in Javascript, with implementation instructions for publishing and receiving DEAs files are open access on GitHub \cite{zheng2020github}. 

Based on this prototype, experiments were conducted covering the complete process of publishing and receiving DEAs files, i.e. the vehicle company's requirement models and the supply agencies' solution models. These models were developed using an MBSE tool \textit{MetaGraph} \cite{Lu2020} (http://www.zkhoneycomb.com/).

\begin{figure*}[!t]
\centering
\includegraphics[width=7.0in]{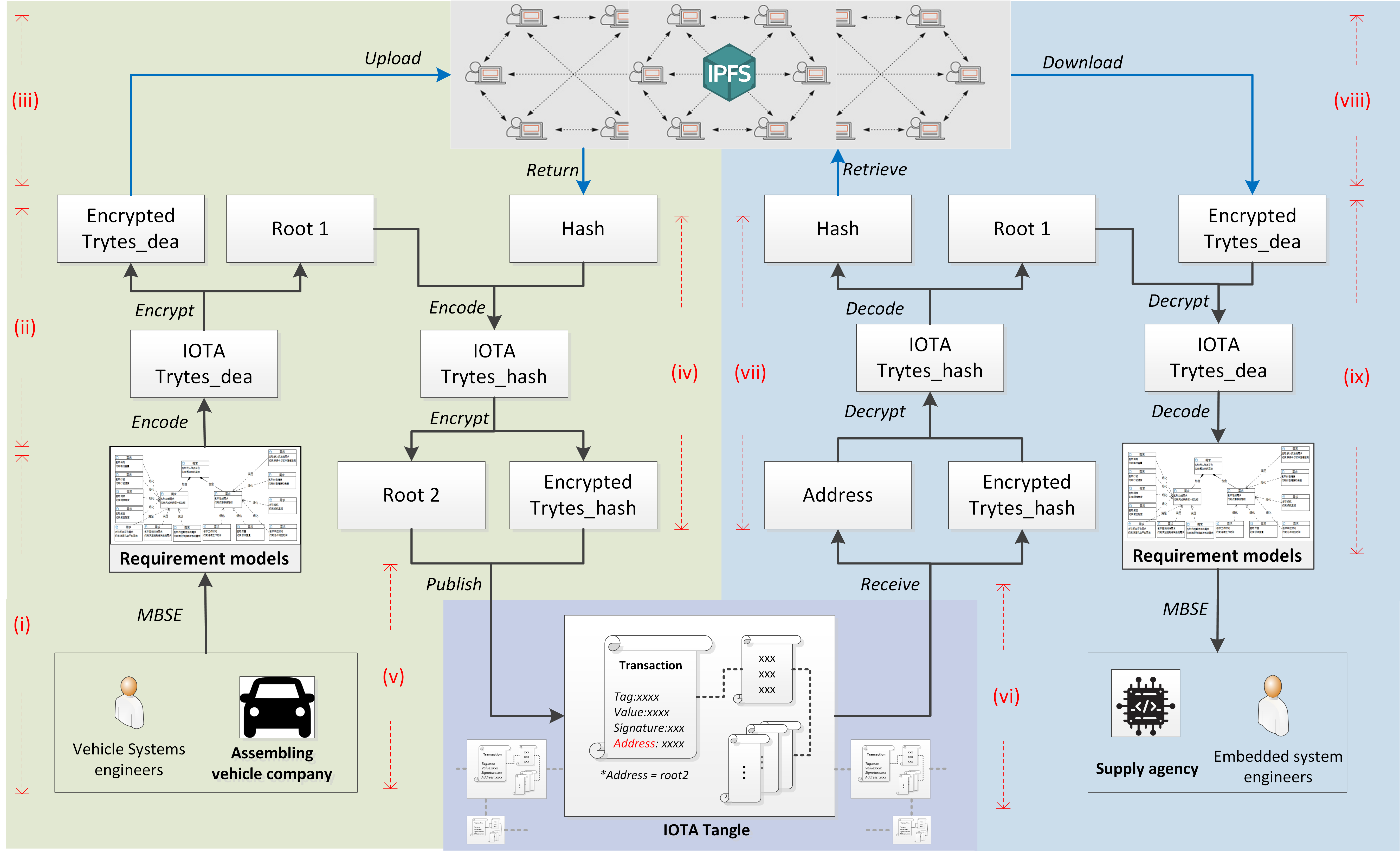}
\caption{DEAs Marketplace workflow}
\label{F:casestudy}
\end{figure*}

\begin{figure*}[!t]
\centering
\includegraphics[width=6.0in]{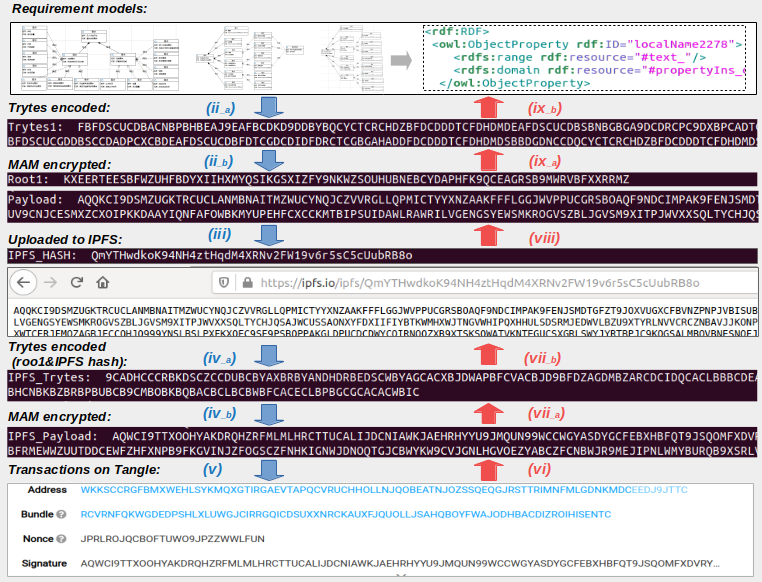}
\caption{Experiment of publishing and receiving requirement models through DEA marketplace}
\label{F:experiment}
\end{figure*}
\subsubsection{Experiment}
The detailed workflow of publishing and receiving the requirement models with public mode through the marketplace is shown in Figure \ref{F:casestudy}. The screenshots of the output from each step based on an example is presented in Figure \ref{F:experiment}. It can be divided into the following steps.

\renewcommand{\labelenumi}{(\roman{enumi})}
\begin{enumerate}
    \item \textbf{Preparing DEAs:} The vehicle systems engineers develop requirement models, using the GOPPRR approach and export them as OWL files \cite{10.1007/978-3-030-16181-1_40} making them ready for transferring.
    \item \textbf{Encoding and encryption:} In order to protect the privacy of the DEAs on the IPFS network, they should be encrypted locally before uploading to IPFS. We used the encryption mechanism provided by IOTA, The original files are first encoded into trytes (\textit{Trytes1: FBFDSCUC...}), which is the data encoding format of IOTA based on ternary numeral system \cite{popov2016tangle}. Then the trytes data are encrypted with the MAM signature scheme, producing the encrypted data (\textit{Payload: AQQKCI9D...}) and a MHT root (\textit{Root1: KXEERTEE...}), which is the decryption key.
    \item \textbf{Uploading to IPFS:} The encrypted data are uploaded to the IPFS network and a content identification (CID) hash code (\textit{Payload: QmYTHwdk...}) will be generated which can be used to retrieve the uploaded content (\textit{AQQKCI9D...}) from IPFS network (\textit{https://ipfs.io/ipfs/QmYTHwdk...}).
    \item \textbf{IPFS CID  encryption:} Using the same encoding and encryption approach, the IPFS hash and the root of the previous encryption are encoded (\textit{IPFS\_Trytes: 9CADHCCC...}) and encrypted to construct the payload (\textit{IPFS\_Payload: AQWCI9TT...}) of the IOTA transaction. A second root (\textit{Root: WKKSCCRG...}) will be generated which can be used to decrypt the payload.
    \item \textbf{Publishing payload:} Using the MAM API, the payload will be published to the IOTA Tangle and one or more transactions will be generated. In public mode the second root (\textit{WKKSCCRG...}) will be used as the address of the transaction. Certain tags and descriptions can also be added to the transactions to facilitate future search or subscription.
    \item \textbf{Receiving transactions:} Once the IOTA transactions are approved, the suppliers can retrieve them through the transaction address. Certain searching service could be added to facilitate the transaction finding process.
    \item \textbf{Payload decryption:} After received the transaction, the supplier can decrypt the payload using the address, which is the same as the root in public mode, and then decode the trytes format content to obtain the CID hash code and the \textit{root1}. 
    \item \textbf{Download from IPFS:} Using the CID, the encrypted content can be downloaded from IPFS network.
    \item \textbf{DEAs Decryption:} The downloaded content can be decrypted with \textit{root1} to get the trytes format content. After decoding, the original requirement models will be readable to the embedded systems engineers of the supplier.
\end{enumerate}
After analyzing the received requirement models, the suppliers will develop corresponding solutions and prepare the solution models with the same MBSE approach. The solution models will be published to the marketplace following a similar workflow. In this case, they can use restricted mode instead of public mode. The only difference is that, the address of the transactions on IOTA Tangle is the hash of the root and a extra side key is required to decrypt the transaction payload. Therefore, the suppliers can share their side keys only to the vehicle company to authorize them with access to the solution models.

\subsubsection{Performance evaluation}
In this study, we used a public IOTA node (https://www.iotaqubic.us:443) combined with a local IPFS node to build the experiment environment. A series of experiments were conducted to evaluate the performance of the prototype. Figure \ref{F:experiment} presents some output screenshots during publishing a requirement model which correspond to the workflow shown in Figure \ref{F:casestudy}. Three groups of MBSE models were created and each of them was published to the IOTA Tangle 10 times. During each iteration, the time of three procedures were measured, i.e. local processing, attaching transactions to the Tangle and transactions been confirmed. The local processing includes data encoding, encryption and uploading to IPFS. The results are listed in Table \ref{table:case} and the original experiment records are also open access \cite{zheng2020github}.

According to the experiment results, the local processing time show positive correlation with the size of the DEAs files. This agrees with the expectation that the encoding, encrypting and uploading time increases when the data size grows. The time of attaching transactions to the Tangle is not impacted by the file size because only the encrypted CID, with fixed length, is published to IOTA Tangle. However, it is worth to note that the attaching time varies depending on the conditions of the chosen IOTA node, such as computing power, load, number of neighbours. The confirming time shows high variance which is due to the tip selection algorithm of IOTA protocol \cite{popov2016tangle}.  

\begin{table*}[]
\begin{center}
\caption{Performance of the marketplace prototype}
\label{table:case}
\begin{tabular}{llccc}\toprule
\multirow{2}{*}{DEAs files} & \multirow{2}{*}{Size} & \multicolumn{3}{c}{Time (seconds): mean \textit{(SD, min, max)}}  \\ \cline{3-5}
  &  & Processing & Attaching & Confirming \\\hline
  Case1 models &2.9Mb  & 0.80 \textit{(0.07, 0.72, 0.97)} & 12.63 \textit{(10.82, 4.70, 33.94)} & 158.90 \textit{(190.08, 46.00, 677.00)} \\
  Case2 models &4.6Mb  & 1.21 \textit{(0.12, 1.09, 1.39)} & 16.21 \textit{(18.69, 4.40, 48.03)} & 187.70 \textit{(175.00, 42.00, 663.00)} \\
  Case3 models &5.6Mb  & 1.37 \textit{(0.09, 1.29, 1.53)} & 18.33 \textit{(21.74, 4.47, 52.57)} & 172.90 \textit{(178.80, 37.00, 657.00)}  \\ \hline
\end{tabular}
\end{center}
\end{table*}


\subsubsection{Privacy and security analysis}
The privacy and security analysis for the proposed marketplace is provided as follows:  
\begin{itemize}
    \item Authentication: The proposed marketplace supports two different privacy modes depending on the publisher's needs. Data published to the marketplace in public mode are accessible to all stakeholders without restriction. In contrast, data published in restricted mode are only open to those authorized with an extra encryption key. When IPFS is used, the data uploaded to IPFS are encrypted in both modes, which keeps the data only open to the marketplace stakeholders. Even the IPFS content is compromised, the attacker will not get access to the data as they are encrypted before uploading.
    \item Nontampering: The tangle-based consensus protocol enables the published transactions on the marketplace with tamper-resistance characteristic. Once the transaction is approved, nobody will be capable of modifying unless the entire network is compromised. The data uploaded to IPFS network are also tamper-resistance owing to the hash function provided by IPFS. Any modification to the original data will result in a totally different identification. Besides, the encryption procedure, which also uses hash function, before uploading to IPFS adds an extra layer of protection.
    \item Decentralization: The proposed marketplace is based on a public tangle-based DLT protocol. This protocol was designed to be fully decentralized because no miners are required. Therefore the centralization risk caused by mining pools is eliminated. 
\end{itemize}

\subsubsection{Limitations}
The enabling technologies, especially the adopted DLT solution, for the proposed marketplace are promising but still evolving. Some limitations remain to be addressed. 

The efficiency and security of the adopted IOTA Tangle DLT protocol heavily relies on the size of the network, i.e. the number of connect nodes and participants. The nodes connected and more transactions submitted, the network will have the higher security and transaction rate. Currently, the IOTA Tangle is still in its early phase with limited number of nodes. To protect the network from malicious attacks a temporary mechanism called the \textit{Coordinator} has been used to directly or indirectly validate the transactions. This mechanism brings the concerns of centralization and thus the single point of failure risk. This limitation is expected to be resolved in the near future by launching the \textit{Coordicide} modular \cite{popov2020coordicide}, which will remove the \textit{Coordinator} and make the Tangle fully decentralized. 

Pricing strategy and payment method are essential components for a mature marketplace which are not investigated in this study. The main reason is that during systems development, the price is usually offered by the DEAs providers. This is different from IoT data trading or smart energy trading etc., where the price is usually dynamically defined by the marketplace. However, certain price comparison and recommendation mechanisms will be beneficial and could be future work of this study.

\section{Conclusion}
This paper proposes a decentralized open marketplace to facilitate DEAs exchange during system developments among inter-organization stakeholders.
Following the systems thinking methodology, the functional requirements of the marketplace are defined based on scenario analysis. A DEAs marketplace framework is designed and its key concepts are formally defined. The GOPPRR approach is adopted to create DEAs, thus to enable the interoperability of the MBSE models. Moreover, the tangle-based DLT solution, IOTA Tangle and its MAM messaging protocol, is adopted to enable secure, nontampering, low-cost and decentralized DEAs exchange. A proof-of-concept implementation of the marketplace is conducted based on a case study. Experiment results prove the feasibility of the proposed approach. 


%

\section*{Acknowledgment}

The work presented in this paper is supported by the EU H2020 project (869951) FACTLOG-Energy-aware Factory Analytics for Process Industries and EU H2020 project (825030) QU4LITY Digital Reality in Zero Defect Manufacturing and the InnoSwiss IMPULSE project on Digital Twins.

\ifCLASSOPTIONcaptionsoff
  \newpage
\fi

\bibliographystyle{IEEEtran}
\bibliography{MODELSBIB}


\begin{IEEEbiography}[{\includegraphics[width=1in,height=1.25in,clip,keepaspectratio]{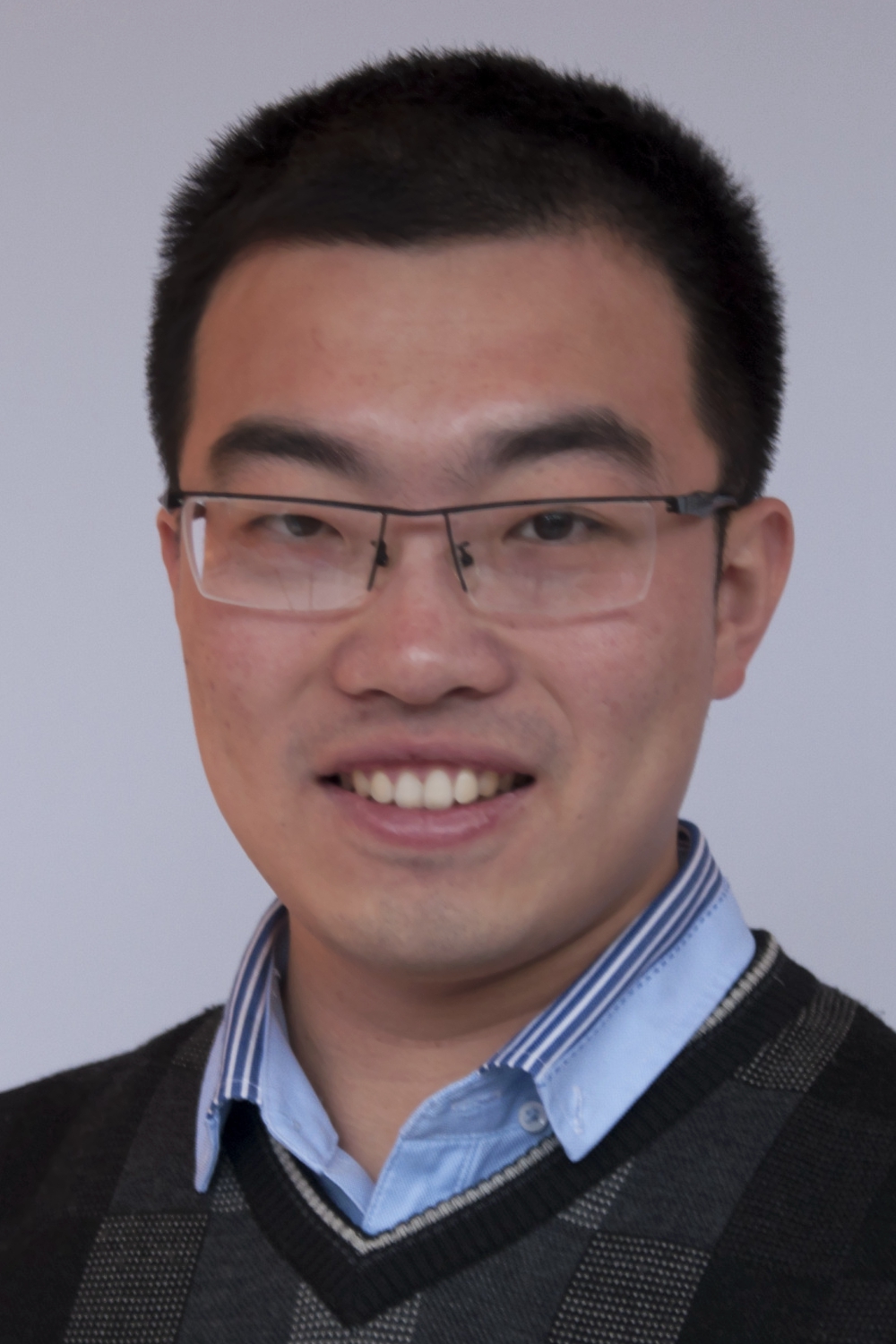}}]{Jinzhi Lu}
, CSEP, is a research scientist at EPFL. He got his ph.d degree at KTH Royal Institute of Technology, Mechatronics Division in 2019. His research interest is MBSE tool-chain design and MBSE enterprise transitioning. He is senior member of China Council on Systems Engineering (CCOSE), China Council on Systems Engineering.
\end{IEEEbiography}

\begin{IEEEbiography}[{\includegraphics[width=1in,height=1.25in,clip,keepaspectratio]{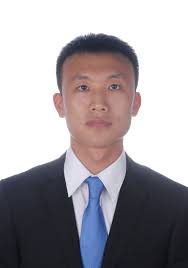}}]
{Xiaochen Zheng}
Ph.D, received his doctoral degree from Universidad Politécnica de Madrid. Before that he studied in Shandong University in Mechanical Engineering and obtained his bachelor and master degree. He is now working at École Polytechnique Fédérale de Lausanne as  a postdoctoral scientist. His research interests include Internet of Things, Machine learning, Wearable technology, Distributed ledger technology and their applications in industry and healthcare etc.
\end{IEEEbiography}

\begin{IEEEbiography}
[{\includegraphics[width=1in,height=1.25in,clip,keepaspectratio]{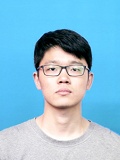}}]
{Zhenchao Hu}
is a PhD student at SJTU. He received his bachelor degree at Wuhan University in Mechanical and Electronic Engineering. His research interest remain in the data integration, prognostics health management, etc.
\end{IEEEbiography}

\begin{IEEEbiography}
[{\includegraphics[width=1in,height=1.25in,clip,keepaspectratio]{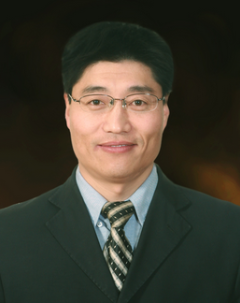}}]
{Huisheng Zhang}
Zhang was promoted as an associate professor at SJTU in 2001. In 2005-2006, he was selected to be the research scientist in the Department of Safety and Nuclear Engineering, AREVA NP in France by FFCSA. In December 2009, he was promoted as a full professor at SJTU. Prof. Zhang’s research interests remain in the modeling and simulation, dynamics and control of power plant, health management system etc. 
\end{IEEEbiography}

\begin{IEEEbiography}[{\includegraphics[width=1in,height=1.25in,clip,keepaspectratio]{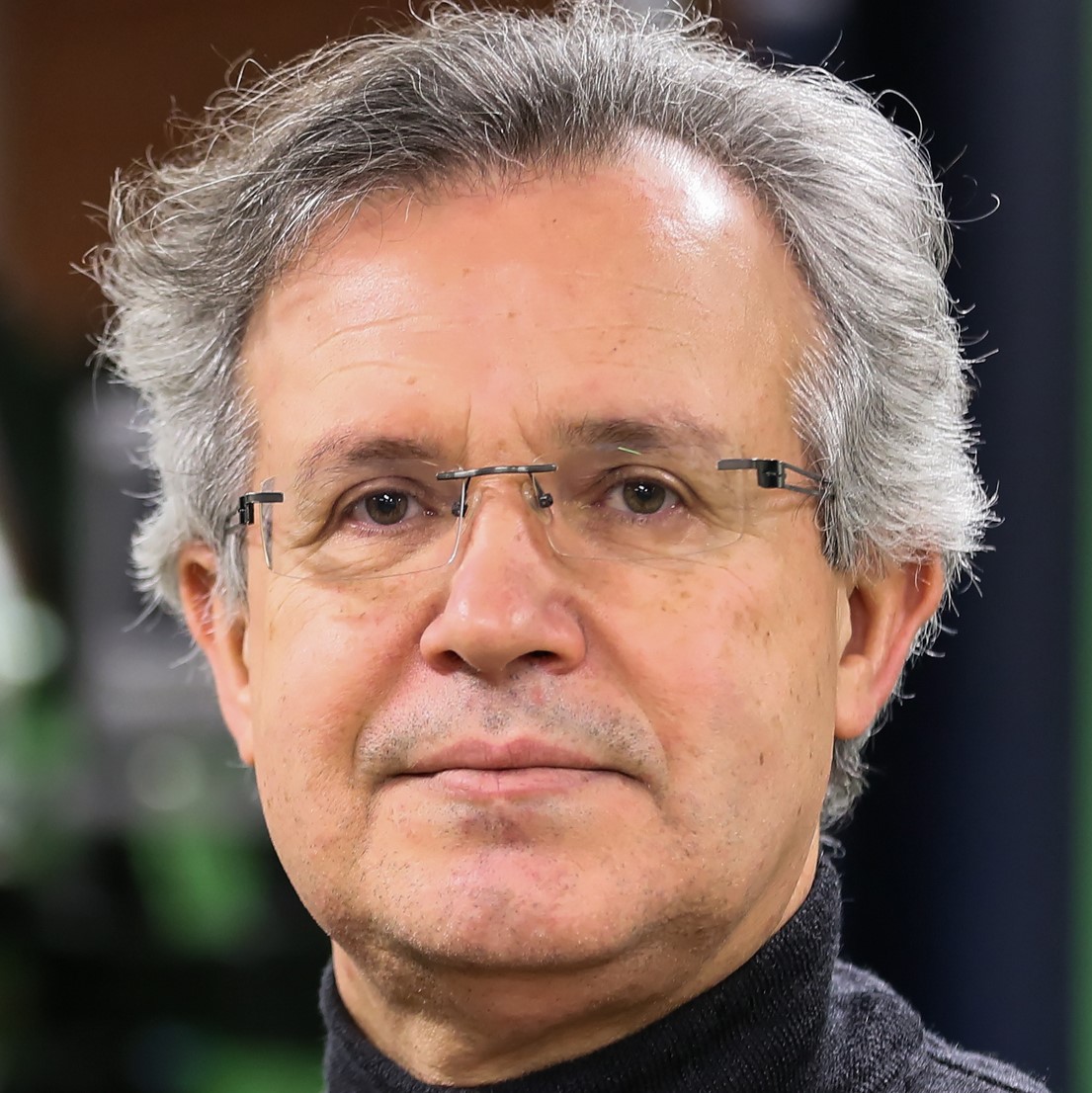}}]
{Dimitris Kiritsis}
is Faculty Member at the Institute of Mechanical Engineering of the School of Engineering of EPFL, Switzerland, where he is leading a research group on ICT for Sustainable Manufacturing. He serves also as Director of the doctoral Program of EPFL on Robotics, Control and Intelligent Systems (EDRS). His research interests are Closed Loop Lifecycle Management, IoT, Semantic Technologies and Data Analytics for Engineering Applications. 
\end{IEEEbiography}

\end{document}